\documentclass[aps,prl,reprint,amssymb,showpacs,onecolumn,superscriptaddress,notitlepage]{revtex4-1}
\usepackage{bm}
\usepackage{graphicx}
\usepackage{dcolumn}
\usepackage{braket}
\usepackage{natbib}
\usepackage{amsmath}
\usepackage{color}
\usepackage{ulem}
\usepackage{siunitx}

\definecolor{gray}{rgb}{0.7,0.7,0.7}
\definecolor{orange}{rgb}{1, 0.4, 0}
\definecolor{dgreen}{rgb}{0.0, 0.4, 0.0}
\definecolor{yblue}{rgb}{0.06, 0.3, 0.57}

\newcommand\ldsout{\bgroup\markoverwith{\textcolor{blue}{\rule[0.5ex]{2pt}{0.4pt}}}\ULon}

\begin{document}

\title{Quantum hydrodynamics of a single particle}
\author{D. G. Su\'arez-Forero}
\affiliation{CNR NANOTEC, Institute of Nanotechnology, Via Monteroni, 73100 Lecce, Italy}
\affiliation{Dipartimento di Ingegneria dell'Innovazione, Universit\`{a} del Salento, via per Monteroni, km 1, 73100 Lecce, Italy’}
\author{V. Ardizzone}
\email{v.ardizzone85@gmail.com}
\affiliation{CNR NANOTEC, Institute of Nanotechnology, Via Monteroni, 73100 Lecce, Italy}
\author{S. F. Covre da Silva}
\affiliation{Institute of Semiconductor and Solid State Physics,
Johannes Kepler University, Linz, Altenbergerstr. 69, 4040, Austria}
\author{M. Reindl}
\affiliation{Institute of Semiconductor and Solid State Physics,
Johannes Kepler University, Linz, Altenbergerstr. 69, 4040, Austria}
\author{A. Fieramosca}
\affiliation{CNR NANOTEC, Institute of Nanotechnology, Via Monteroni, 73100 Lecce, Italy}
\affiliation{Dipartimento di Fisica, Universit\`{a} del Salento, Strada Provinciale
Lecce-Monteroni, Campus Ecotekne, Lecce 73100, Italy}
\author{L. Polimeno}
\affiliation{CNR NANOTEC, Institute of Nanotechnology, Via Monteroni, 73100 Lecce, Italy}
\affiliation{Dipartimento di Fisica, Universit\`{a} del Salento, Strada Provinciale
Lecce-Monteroni, Campus Ecotekne, Lecce 73100, Italy}
\author{M. de Giorgi}
\affiliation{CNR NANOTEC, Institute of Nanotechnology, Via Monteroni, 73100 Lecce, Italy}
\author{L. Dominici}
\affiliation{CNR NANOTEC, Institute of Nanotechnology, Via Monteroni, 73100 Lecce, Italy}
\author{L. N. Pfeiffer}
\affiliation{PRISM, Princeton Institute for the Science and Technology of Materials, Princeton Unviversity, Princeton, NJ 08540}
\author{G. Gigli}
\affiliation{Dipartimento di Fisica, Universit\`{a} del Salento, Strada Provinciale
Lecce-Monteroni, Campus Ecotekne, Lecce 73100, Italy}
\author{D. Ballarini}
\affiliation{CNR NANOTEC, Institute of Nanotechnology, Via Monteroni, 73100 Lecce, Italy}
\author{F. Laussy}
\affiliation{Faculty of Science and Engineering, University of Wolverhampton, Wulfruna Street, Wolverhampton WV1 1LY, UK.}
\affiliation{Russian Quantum Center, Novaya 100, 143025 Skolkovo, Moscow Region, Russia}
\author{A. Rastelli}
\affiliation{Institute of Semiconductor and Solid State Physics,
Johannes Kepler University, Linz, Altenbergerstr. 69, 4040, Austria}
\author{D. Sanvitto}
\email{daniele.sanvitto@nanotec.cnr.it}
\affiliation{CNR NANOTEC, Institute of Nanotechnology, Via Monteroni, 73100 Lecce, Italy}


\begin{abstract}
  Semiconductor devices are strong competitors in the race for the
  development of quantum computational systems. In this work, we
  interface two semiconductor building blocks of different
  dimensionality and with complementary properties: (1) a quantum dot
  hosting a single exciton and acting as a nearly ideal single-photon
  emitter and (2) a quantum well in a 2D microcavity sustaining
  polaritons, which are known for their strong interactions and unique hydrodynamics properties including ultrafast real-time monitoring of their propagation and phase-mapping. In the present experiment we
  can thus observe how the injected single particles propagate and evolve inside the microcavity, giving rise to hydrodynamics features typical of macroscopic systems despite their intrinsic genuine quantum nature.  In the presence of a structural defect, we observe
  the celebrated quantum interference of a single particle that
  produces fringes reminiscent of a wave propagation. While this behaviour could be theoretically expected, our imaging of
  such an interference pattern, together with a measurement of
  antibunching, constitutes the first demonstration of spatial mapping of the self-interference of a single quantum particle hitting an obstacle.
\end{abstract}

\maketitle
\section{Introduction}

The generation, manipulation and detection of
on-chip single photons is key to the development of photonic-based
quantum information technologies~\cite{Won2019}. Integrated optics
(IO) devices working in the single-particle regime will enable the
deployment of quantum information processing for both fundamental
research and technological applications. IO chips should provide
qubits generation, processing and readout. For instance, qubits
generation can be implemented by semiconductor quantum dots (QDs)
\cite{Huber2018a,Elshaari2017} or parametric sources~\cite{PhysRevA.60.R773,Fedrizzi:07}. Superconducting
single-photon detectors~\cite{You2017} seem to be among the most
promising candidates to date for integrated qubit detection
\cite{Gourgues2019,Schwartz2018a}. Most of the optical circuits used for quantum
information 
so far rely on linear properties of single photons propagation
\cite{Carolan2015} or on optical non-linearities of $\chi^2$ or
$\chi^3$ materials~\cite{Luo2019, Moss2013, Jin2014}. By combining
these elements, several functionalities such as quantum logic gates
\cite{Crespi2011, Politi2008}, boson sampling~\cite{Tillmann2013},
quantum interference or quantum metrology~\cite{Giovannetti2011} have
been demonstrated. However, present schemes for single qubits
manipulation face real challenges, relying on complex cascades of linear optical elements or on weak nonlinear susceptibilities requiring long interaction paths. These features could result in severe limitations in
the scalability and miniaturization of future devices.

Microcavity polaritons, the hybrid light-matter quasiparticles emerging
from the strong coupling between a cavity mode and an excitonic
transition, could represent a promising alternative to achieve quantum
information processing in integrated optical circuits
\cite{Sanvitto2016}.  Their intrinsically interacting nature, inherited
from their excitonic component, together with their long coherence
time, inherited from their light component, make them strong
candidates to perform nonlinear logic operations without losing
information~\cite{Ballarini2013, Leyder2007}. While collective
mesoscopic phenomena involving microcavity polaritons---such as
polariton lasing, superfluidity or optical parametric
oscillation---have been extensively studied~\cite{Carusotto2013},
there has been little exploration of their quantum, few-particle limit,
i.e., involving non-Gaussian polariton states. The experimental
demonstrations of polaritonic quantum behaviour have been mainly
limited to Gaussian mixtures at best~\cite{Sassermann2018,
  Boulier2014, Ardizzone2012}, with only recent progress towards the
generation of non-Gaussian states, following the demonstration of
polariton blockade~\cite{Delteil2019,munoz2019,Gerace2019}.  Taking
 the entirely different approach of exciting polaritons with quantum
light~\cite{lopezcarreno15a}, it was demonstrated experimentally that
the creation and recombination of polaritons in a semiconductor
microcavity can be done without damaging the quantum coherence of
nonclassical states, which is a necessary condition for any quantum
information processing~\cite{Cuevas2018}.

In this work, we demonstrate how one can use single photons emitted by an external semiconductor QD to generate, inject and propagate
individual microcavity polaritons - a fundamental milestone for the developement of future polariton quantum devices. Moreover, with this experiment we show to be able to map the propagation of single polaritons in the two dimensional space during the propagation time, an thus bring down to its ultimate single-particle limit, the prominent and remarkable propagation of a polariton fluid.  This is the first step towards several single-particle configurations in a solid-state,
integrable, setup manipulating highly
interacting single-polariton qubits.  In our case, still with a single polariton, by imaging its propagation across an obstacle acting as a scattering
center, we are able to observe the wavelike interferences that are
produced by what is otherwise one polariton alone.  This
observation represents  an alternative version of the
double-slit experiment, directly demonstrating the wave-particle duality
for individual microcavity polaritons \cite{Kolenderski2014, Feynman2010}. While wave-particle duality is somehow expected for quantum single particles, our experiment represents the first 2D mapping of such a behaviour. In our case, instead of a double slit or a single particle splitting, we observe the interference effect on a multiple path-scattering propagation of a single particle flying against an obstacle smaller than the wavepacket size, providing a direct imaging of the wave-particle duality of these light-matter excitations \cite{Moreau2019}.

\section{Results}

A scheme of the experimental setup is shown in Fig.~\ref{setup}a. It is composed of three main parts: i) the generation of single photons, ii) the injection and propagation of single polaritons and iii) detection. The imaging and spectroscopy experiments were performed in both reflection and transmission configurations. The latter required a processing of the substrate to enable the transmission, and an optimization of the signal intensity by increasing the single-photon emission rate. A more detailed representation of the experimental setup can be found in the Supplementary Material. Despite its conceptual simplicity, our hybrid approach that couples a single photon source, providing the qubits, to polaritons in a high quality factor microcavity, propagating as single particles, presents considerable technical difficulties. First of all, the use of a tunable source of heralded single photons, i.~e.~ Spontaneous Parametric Down Conversion (SPDC) in non-linear crystals, must be ruled out. This kind of source would imply a heralded measurement in order to be sure to measure the propagation of single photons, resulting in a very cumbersome implementation with common imaging systems. Additionally, the broad emission spectra of SPDC sources~\cite{Fedrizzi2007} would be poorly coupled to the microcavity, whose high quality factor is necessary to confine the quantum state long enough to sustain polariton propagation. We decided then to use single QDs as a deterministic single photon source with an emission linewidth compatible with the narrow polariton resonance. However, most common QD systems (InGaAs QDs grown with the  Stranski-Krastanov method) have a typical emission range incompatible with GaAs/AlGaAs microcavities. This makes necessary the use of GaAs QDs produced by Al droplet etching, that recently has shown to be capable of producing near ideal photon sources comparable with best commercial single photon emitters~\cite{Schweickert2018}, showing values of $g^{(2)}(\tau=0)$ below $10^{-4}$. The choice of QDs as single photon sources entails an additional obstacle to overcome: the necessity of keeping both systems, QD  and microcavity, at cryogenic temperatures.
Here we demonstrate that these issues can be overcome obtaining an alternative platform for the study of polariton systems in the single or few particle regime involving several qubits.

\begin{figure*}
 \begin{center}
  \includegraphics[width=0.99\linewidth]{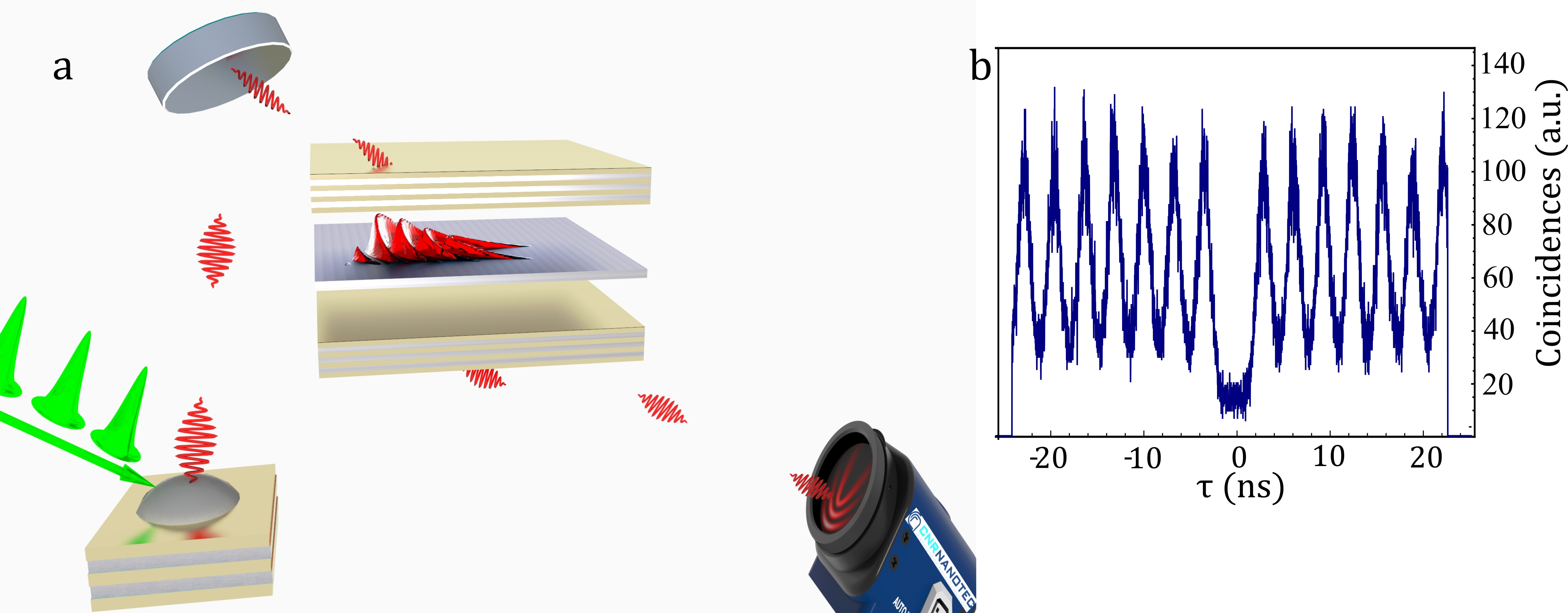}
  \caption{a) Schematic picture of the experiment: a pulsed laser pumps a QD to generate single photons that are injected inside a semiconductor microcavity. The image of the single polariton propagation is acquired in an EMCCD. b) Second-order correlation
    function of a QD emission when multiplexing the pump
    pulse-rate to 320 MHz. The antibunching value of
    $g^{(2)}(0)=0.16\pm0.05$ is an unequivocal signature of single photon emission from these QDs. \label{setup}}
 \end{center} 
\end{figure*}

\begin{figure}
 \begin{center}
  \includegraphics[width=0.6\columnwidth]{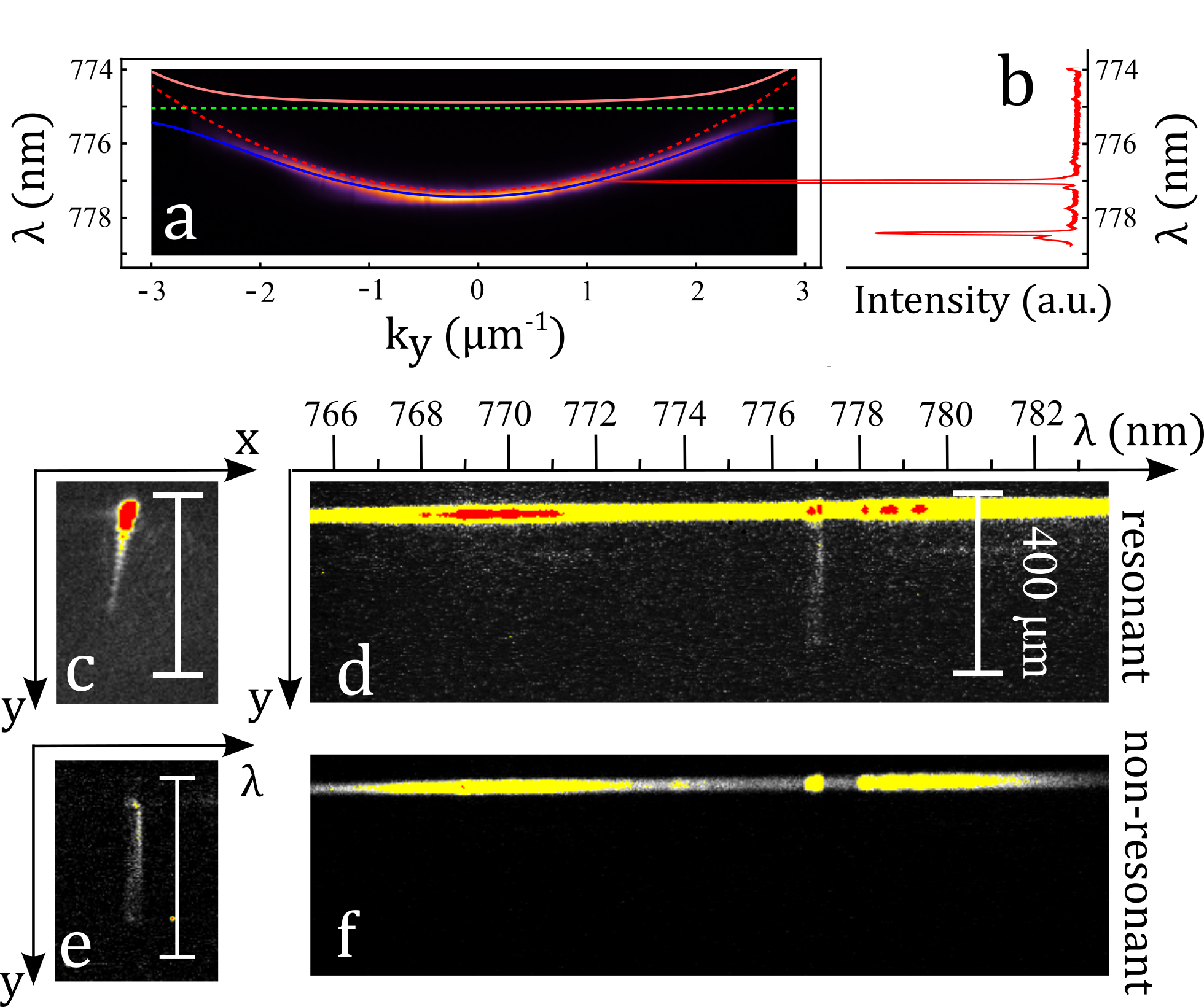}
  \caption{a) Energy dispersion of the microcavity-quantum well system
    at the point of incidence of photons in reflection configuration, compared with the emission
    spectrum of the pumping QD, shown in panel b. At $k_y\approx1.1$
    $\mu$m$^{-1}$ the exciton's energy is in resonance with the LPB, allowing a resonant polariton injection into the microcavity. Polaritons injected with this in-plane momentum propagate with a group velocity $v_g\sim2.1$ $\mu m/ps$, as deduced in the Supplementary Material. c)
    real space image of the single polariton propagation. d) resolving
    in energy the propagation in c, it can be evidenced how only the
    QD exciton peak couples into the system and propagates. e) to have
    a better image of the propagation, the non-coupled reflected light
    is blocked using a spatial filter. Polariton propagation distances
    are measured up to $\sim400$ $\mu$m. f) moving at different
    microcavity-quantum well detuning, the QD exciton is not on
    resonance with the polariton dispersion and, indeed, no
    propagation is evidenced in this situation.}\label{prop_refl}
 \end{center}
\end{figure}

In the following experiments, single photons are generated from GaAs QDs
fabricated by Al droplet etching and embedded in a low-Q cavity
consisting of a $\lambda/2$ layer of Al$_{0.4}$Ga$_{0.6}$As with QDs
sandwiched between two Bragg mirrors made of 9 and 2 pairs of
Al$_{0.95}$Ga$_{0.05}$As (\SI{67}{\nano\meter} thick) and
Al$_{0.2}$Ga$_{0.8}$As (\SI{55}{\nano\meter}
thick)~\cite{Schweickert2018, Huber2018}. The QDs are pumped with a
pulsed laser either at $\lambda=\SI{415}{\nano\meter}$ with a pulse
duration of $\sim\SI{30}{\femto\second}$ or at $\SI{760}{\nano\meter}$
with a pulse duration of $\sim\SI{5}{\pico\second}$. Thanks to a pair
of cascaded Michelson and Morley interferometers, we can increase the
laser repetition rate by a factor of four, from $80$ MHz up to $320$
MHz, in order to increase the number of photons per second (for more
details on the photon-rate quadruplication, see the Supplementary
Material). The QD size was optimized to obtain emission wavelengths around $\SI{775}{\nano\meter}$. The generated photons are coupled to a single mode fiber and then used to pump the polaritonic device, a $\lambda/2$ microcavity, which is made of two Bragg mirrors of 40 and 32 pairs of Al$_{0.96}$Ga$_{0.04}$As ($\SI{67}{\nano\meter}$ thick) and Al$_{0.2}$Ga$_{0.8}$As ($\SI{55}{\nano\meter}$ thick) with a quantum well $\SI{7.2}{\nano\meter}$ wide embedded in the center. Both
systems, QDs and the microcavity sample, are cooled down to cryogenic
temperatures of $\SI{3.8}{\kelvin}$ and $\SI{8.5}{\kelvin}$,
respectively.  A comparison between the spectra of the microcavity and
the quantum dot is shown in Fig.~\ref{prop_refl}a, b, respectively, despite the mentioned technical difficult, we succeeded to grow QDs with a transition energy precisely in the energy range covered by theLower Polariton Branch (LPB) of the microcavity-quantum well system~\cite{Kavokin2008}. 
By carefully tuning the single-photons
injection angle, the QD emission can be resonantly coupled to the
LPB. The quantum nature of the light is tested by measuring the second-order correlation function
$g^{(2)}(\tau = 0)$ with a Hanbury Brown and Twiss setup (HBT,
Fig.~\ref{setup}b), finding a value of $g^{(2)}(0)=0.16\pm0.05$. The small residual difference from zero is attributed to
the non-resonant excitation as well as the slow carrier relaxation in
the QDs~\cite{Huber2017}. 
The real space images have been detected by an Enhanced Charged Coupled Device (EMCCD)
camera coupled to a monochromator to allow energy resolved
measurements.

Figure \ref{prop_refl}c shows the real-space image of the microcavity
surface under single-photon excitation. The image shows two main
features: a bright saturated circular spot and a weaker elongated
spot. We assign the former to the reflected uncoupled single photons
and the latter to the resonantly excited single polaritons that
propagate inside the microcavity due to their externally imparted
in-plane momentum. To prove this, we acquire an energy-resolved image
(see Fig.~\ref{prop_refl}d). This image shows that,
from all the peaks in the QD
  emission, the neutral exciton peak resonant with the LPB is the
only one able to couple and propagate inside the sample, as evidenced by the faint but continuous vertical trace at the corresponding wavelength. The high Q factor of the
cavity results in high energy selectivity and long propagation
distances, close to $\SI{400}{\micro\meter}$.  To further prove that
the propagation corresponds to a resonant polariton injection, the
sample is moved to a point in which the LPB is not resonant with the
QD emission, as shown in Fig.~\ref{prop_refl}f. In this case, no
single-polariton propagation is observed. By fitting the intensity
profile of the propagating polaritons with an exponential decay and
considering that single polaritons are injected with an in-plane
momentum of $\SI{1.1}{\per\micro\meter}$ (see Supplementary Material),
we obtain a polariton lifetime $\tau \sim\SI{25}{\pico\second}$, which is in agreement with the lifetime that can be deduced from the
polariton linewidth.

\begin{figure}
 \begin{center}
  \includegraphics[width=0.6\columnwidth]{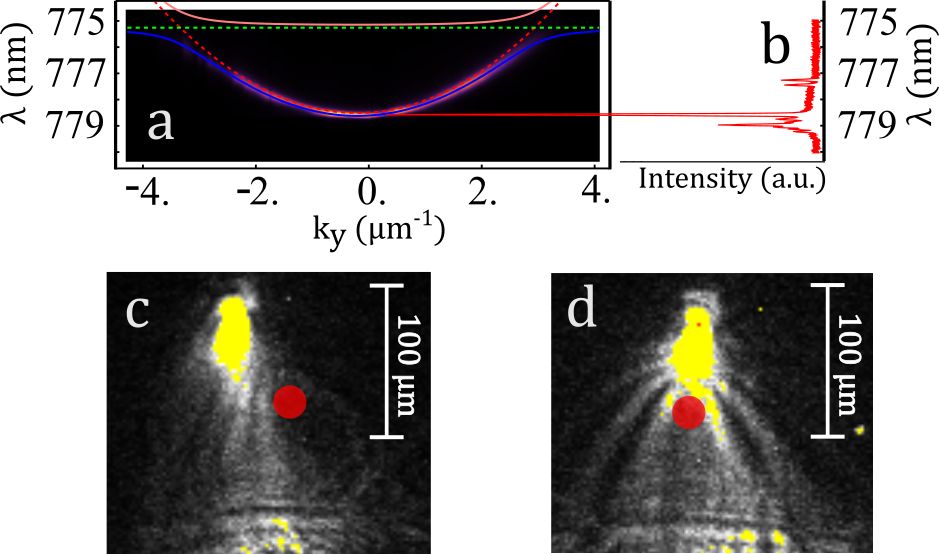}
  \caption{a) Energy dispersion of the microcavity at the photon
    injection point in transmission configuration, compared with the emission spectrum of the
    selected QD, shown in panel b; notice that the exciton's energy of
    the QD coincides with the state of the LPB with in-plane momentum
    $k_y\approx0.28$ $\mu$m$^{-1}$; this allows to resonantly pump the
    microcavity with single photons from the QD. \textbf{The second order correlation function of figure \ref{setup} is obtained precisely for this QD emission spectrum;} c) single polariton
    propagation obtained by matching the polariton dispersion in panel
    a, with the single photons in panel b. d) single polariton
    propagation across a defect naturally occurring in the
    microcavity; an interference pattern appears due to the
    self-interference between the incoming wave function and its
    scattering against the defect. The red circle indicates the position of
    the structural defect.}\label{qds}
 \end{center}
\end{figure}

The reflection configuration is compelling for a proof-of-principle demonstration and as a first attempt to evidence the phenomenon, also allowing us to prove that only the single-photon excitonic peak couples to the microcavity and triggers a propagation inside while others get fully reflected. However, in reflection it is impossible to image the region around the
point of injection. In order to get a more comprehensive picture of the single-polariton propagation, we modify our setup to perform experiments in a transmission
configuration. To do that, the polariton sample was processed by wet
etching to remove the absorbing bulk GaAs substrate in selected
regions, and uncovering the microcavity~\cite{Moon1998}. A comparison
of a QD's emission spectrum and a microcavity LPB dispersion
corresponding to an etched region is shown in Figs.~\ref{qds}a and
\ref{qds}b. The microcavity dispersion (panel a) is measured in
transmission.  Again, a QD with an excitonic peak (in this case a
positively charged exciton) near the LPB is chosen to resonantly
excite single polaritons.  Moreover, to increase the amount of signal,
the pulse repetition rate of the laser exciting the QD was
quadruplicated, up to a final rate of $320$ MHz. \textbf{To carefully check the single photon regime we measured the second order correlation function after each doubling of the pump repetition rate. The corresponding measurements, shown in figure 6 of the Supplementary Material, confirm that the QD always behaves as a single photon emitter.} After
this optimization, the microcavity sample is pumped with around $140$k
single photons per second. To decrease the total
recombination time of the exciton, in order to avoid temporal
overlap of the generated photons, the pump wavelength was changed from $\SI{405}{\nano\meter}$ to $\SI{760}{\nano\meter}$. Indeed, the QD decay time involves several processes, but in general, higher pump energies need more non-radiative processes, entailing longer decay times \cite{reindl19a}. Although different QDs were used to pump the cavity in reflection and transmission configurations, in all cases a $g^{(2)}$ measurement such as the one of Fig.~\ref{setup}b was obtained.

Two cases were considered for the single-polariton propagation in
transmission configuration: a ``free propagation'' along the
microcavity (shown in Fig.~\ref{qds}c), and propagation in the
presence of an obstacle (shown in Fig.~\ref{qds}d), given by a
structural defect in the microcavity. Single-polariton propagation in figures \ref{qds} c and d spreads slightly more than in figure \ref{prop_refl} c-e. This is obtained by modifying the excitation beam divergence, making it easier to image defects in the cavity. The polariton group
velocity, estimated through the dispersion relation, is 
$v_g\approx\SI{0.3}{\micro\meter\per\pico\second}$. For a more complete description of the
deduction of the group velocity, we refer to the Supplementary
Material.

\begin{figure}
 \begin{center}
  \includegraphics[width=0.6\columnwidth]{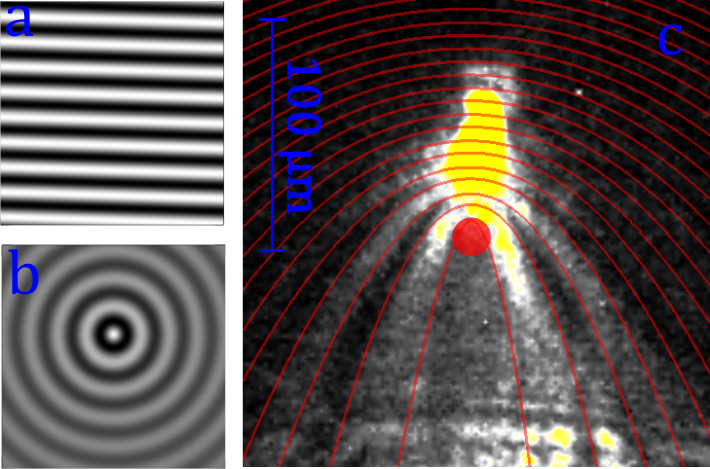}
  \caption{a) Numerical space distribution of the electric field of an
    incoming plane wave and b) for a circular wave, as it could be
    modeled for the light scattered from a punctual structural defect
    in the microcavity.  c) experimental density map from the
    Fig.~\ref{qds}d with superimposed numerical simulations of the
    single polariton self-interference; red lines are contour lines
    having the same intensity and calculated by making the wavefront in a interfere with the circular wave in b. The numerical simulations are obtained
    by assuming an incoming polariton having an in-plane momentum
    $k\sim 0.28$ $\mu m^{-1}$, as in the experiment.}\label{prop}
 \end{center}
\end{figure}

The second case, shown in Fig.~\ref{qds}d, is even more interesting.
It shows the same single-polariton propagation but now across an
obstacle formed by a structural defect highlighted by a red dot. The defect scatters the incoming single polaritons and substantially modifies the propagation
pattern. Clearly, some interferences are formed as seen in
Fig.~\ref{qds}d. While this would be the expected pattern for a
conventional polariton fluid passing an obstacle, here it must be
borne in mind that in the conditions of our experiment, these fringes
arise from integrating photons emitted by polaritons that each
travelled alone in the microcavity, given that they have been
injected there by a strongly antibunched single-photon source with a
repetition rate to polariton lifetime ratio such that each polariton
is separated from the previous and next one by more than 285 thousands
times its lifetime. \textbf{In other words, interference patterns of Fig.~\ref{qds}d and the propagation of Fig.~\ref{qds}c are obtained by integrating several single polariton propagations. There is no interference between subsequent photons spontaneously emitted by polaritons because only one polariton at a time is present in the microcavity}. It is therefore surprising, from a classical
perspective, that one polariton would simultaneously propagate through several distinct trajectories, as is required to produce destructive interferences.  This is a variation of the famous double-slit experiment, in which the wavelike aspect is more fully manifest thanks to the possibility of mapping the polariton field everywhere in the real plane.
Unlike the plethora of similar experiments performed with a screen at the end of the propagation~\cite{Aspect1987,Tonomura1989,Dheur2016,Aspden2016,Zeilinger1988,Arndt1999}, in our case photons are emitted from polaritons by spontaneous emission, and since
this follows an exponential law, they have the same
probability to be emitted at any time of their propagation. In other words, polaritons can provide a full mapping of their spatial dynamics.  In our case, the interference pattern is simply
explained by the interference between a plane wave, representing the
incoming single-polariton (Fig.~\ref{prop}a), and a spherical wave,
representing the scattered polariton (Fig.~\ref{prop}b). The sum of
these two fields amplitudes provide the interference pattern that we
observe with an excellent quantitative agreement, as shown in
Fig.~\ref{prop}c as red lines on top of the experimental background. \textbf{We note that the experimental interference pattern of Fig.~\ref{prop} is consistent with a point-like defect scattering the incoming single polariton plane wave. By point-like defect, we mean that its cross-section is smaller than the polariton wavelength, thus making the defect shape irrelevant for the effects measured here. In our case, the in-plane wavelength of the moving polariton can be estimated to be $\lambda_{//} \approx 2\pi/k \approx 20 \mu m $. To confirm that the defect considered here is point-like, we simulated several interference patterns corresponding to various defect radius $r$ ranging from $r<<\lambda_{//}$ to $r = \lambda_{//}$, see Fig.~\ref{prop} in the Supplementary Materials. It is evident, from these simulations, that when the defect cross-section becomes relevant, the pattern shows high-order interference features with several phase jumps giving rise to straight discontinuities across the main interference fringes. None of these features is present in our experimental data, supporting the hypothesis of a point-like defect. Moreover, the same model reproduces the data of Fig.~\ref{qds}c in which the defect is away from the main propagation axis of the single polariton, see figure 5 of the Supplemental Material.}

\section{Discussion}

The apparent paradox of the double-slit experiment is nowadays familiar in quantum theory. Feynman called it ``the only mystery'' of quantum mechanics~\cite{Feynman2010}. Still, its implications and deep
underlying meaning do not cease to captivate one's
imagination. The polariton platform could contribute to this foundational and fundamental question.
The wonders of quantum mechanics in our spatially mapped version of the double-slit experiment are even more salient from the fact that the fringes are visible ahead of
the obstacle, meaning that even though the single polariton is not yet supposed to know that an obstacle lays ahead, its plane-wave forward motion amplitude of probability is already interfering with its
scattered-backward spherical motion, thus suppressing the spontaneous
emission from a polariton at any point of destructive interference
even ahead of the obstacle.\\
Even if wave-particle duality for single particles is somehow expected in quantum mechanics, the access to the full spatial profile of the field opens new perspectives for deeper investigations of this key phenomena that, more than any other, questions the nature of physical reality. It makes it for instance an interesting testbed for Wheeler's delayed-choice experiment and
variations of it~\cite{vincent07a}, that question whether an
observation can be conditioned on the past history of the particle. We have touched upon already how, in an interesting twist, the future of the particle is brought into question in our configuration. One could stretch the ``particle'' character of the polariton by imposing
tighter constrains on its wavepacket size and localization and/or
passing it through better designed potentials. 
This platform should also be able to explore the violation of causal time-ordering~\cite{koswami18a} in a quantum polariton switch. Also exciting is the possibility to involve multiple polaritons, and study nonlocal effects. In all cases, with the possibility of a reconstruction of the full polariton-field wavefunction, we expect that our first demonstration of a propagating single-polariton will give access to a wide range of fundamental quantum experiments in integrated optics.\\
In conclusion, we have demonstrated the conversion of single photons
from a semiconductor QD into propagating 2D microcavity polaritons, by
resonant injection in a semiconductor planar microcavity, and we have
observed their propagation in a still minimally controlled
environment, i.e., in presence or absence of an obstacle on its way. The observation of a single polariton propagation makes one step further towards the design and
implementation of several single-polaritons devices. At a fundamental
level, we report the first observation of a polariton fluid that
consists of a single particle and have confirmed its wave-particle
duality by observing fringes that result from wave interferences from
states that each consists of a single polariton. In contrast to the
numerous earlier reports on one of the most important and far-reaching
experiments of Physics---the double-slit experiment---we have been able
to provide the interference pattern in the full region of space where the phenomenon occurs. Further investigations of this phenomenon could allow to better understand the fundamental aspects and ontological meaning of quantum theory at large. From an application point of view, the fact that both the QD and polaritons are based on the same material combination opens the route to fully integrated solutions, where polaritons may mediate the interaction of photonic qubits emitted by QDs.

\section{Methods}
\textbf{Microcavity Sample}: We use a $\lambda/2$ cavity embedded between two DBRs formed by respectively 40 and 32 pairs of $Al_{0.96}Ga_{0.04}As$ (67 nm thick) and $Al_{0.2}Ga_{0.8}As$ (55 nm thick). A 7.2 nm wide GaAs quantum well is placed in the center of the cavity, at the maximum of the electric field. The sample substrate has been partially removed by wet etching to measure polariton propagation in transmission geometry. The wet etching process has been carefully calibrated to selectively attack the substrate and the number of pairs of the bottom DBR is not modified. \\

\textbf{QDs Sample}: GaAs QDs are fabricated by Al droplet etching and embedded in a low-Q cavity formed by a $\lambda/2$ layer embedding the Qds sandwiched between two DBRs formed by respectively 9 and 2 pairs of $Al_{0.95}Ga_{0.05}As$/$Al_{0.2}Ga_{0.8}As$ with thickness of 67 nm and 55 nm.\\

\textbf{Experimental realization:} Both samples are kept at cryogenic temperatures, \textbf{in two different cryostats at} 3.8 K for the QDs and 8.5 K for the microcavity. In the reflection configuration, the QDs sample was pumped with a fs pulsed laser at 405 nm with a repetition rate of 80 MHz. In the transmission configuration, the excitation is done with a 780 nm ps pulsed laser, which has been multiplexed by using a cascade of Michelson and Morley interferometers to obtain a final repetition rate of 320 MHz. The emission from the QD is collected in a single-mode fiber optics and used to pump the microcavity sample in a configuration that allows a fine control of the in-plane linear momentum. For the detection, an image of the propagation plane is reconstructed in an Enhaced Charge Coupled Device (EMCCD).

\section{Acknowledgements}
The authors acknowledge Paolo Cazzato for technical support and the ERC project Elecopter grant number 780757 for financial support. AR acknowledges Y. Huo, R. Trotta, D. Huber, H. Huang, and G. Weihs for fruitful discussions and the Linz Institute of Technology (LIT) through the LIT Secure and Correct Systems Lab funded by the state of Upper Austria, and the Austrian Science Fund (FWF): P 29603 for financial support.

\bibliographystyle{unsrt}
\bibliography{Bibliography.bib}

\newpage
\onecolumngrid

\setcounter{equation}{0}
\setcounter{figure}{0}
\setcounter{table}{0}
\setcounter{page}{1}
\makeatletter
\renewcommand{\theequation}{S\arabic{equation}}
\renewcommand{\thefigure}{S\arabic{figure}}
\pagenumbering{roman}

\noindent
\textbf{\Large Supporting Information}\\

\section{Deduction of the polariton's group velocity}
By using a model of coupled oscilators, it is possible to fit the measured dispersion in order to obtain an analytical expression \cite{Kavokin2008}. From the expression of $\omega(k)$, it is possible to obtain the group velocity as $\frac{d\omega}{dk}$. Fig.~S1 shows the plot of the analytical expression found for the LPB (solid blue), UPB (solid pink), bare exciton (dashed green) and cavity mode (dashed red) for the reflection configuration shown in Fig.~2 of the main text. By changing units of the dispersion, in order to have the angular frequency $\omega$ as function of momentum $k$, and taking the first derivative, the analytical expression for the group velocity is obtained. In this case the in-plane momentum of the incoming photons is found to be $\approx1.1$ $\mu m^{-1}$, that corresponds to a group velocity of $\approx 2.1$ $\mu m/ps$.

\begin{figure}[!htb]
    \centering
    \includegraphics[width=0.6\columnwidth]{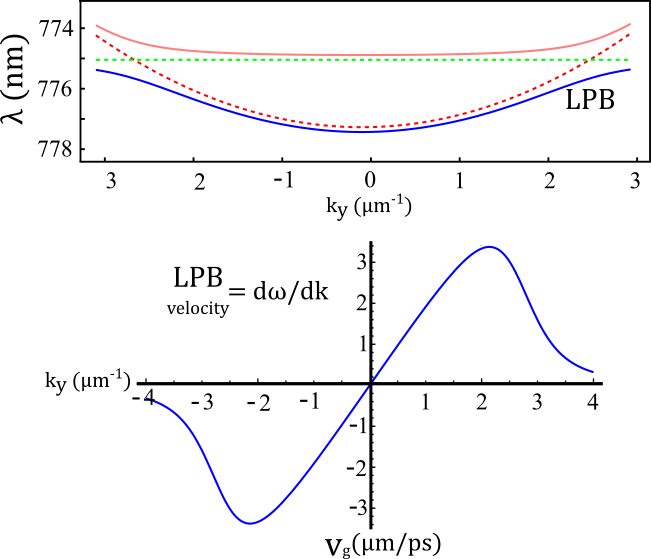}
    \caption{Upper panel: analytical fit of the dispersion shown in figure 2 of the main text. As explained, the QD exciton's energy is in resonance with the LPB at $k_y\approx 1.1$ $\mu m^{-1}$. Lower panel: group velocity for the fitted LPB.}
    \label{fig:my_label}
\end{figure}

\section{Exponential decay and lifetime}
 Panels a and b of figure 2 of the main text show the wavelength and in-plane momentum of the injected particle ($777.0$ nm at $k_y=1.1$ $\mu$m$^{-1}$). As illustrated in last section, from the fitting with a theoretical model, the deduced group velocity is $\approx2.1\mu m/ps$.\\
On the other hand, by making a decay profile of figure 2c of the main text, it is possible to get a decay length by doing an exponential fitting. As shown in figure S2, this length is measured to be $52$ $\mu$m.\\
These two quantities give an estimation of the lifetime of $\tau_t=\textit{l}_{decay}/v_g\approx 25$ ps, a value in agreement with the lifetime measured from the dispersion linewidth.\\

\begin{figure}[!htb]
    \centering
    \includegraphics[width=0.5\columnwidth]{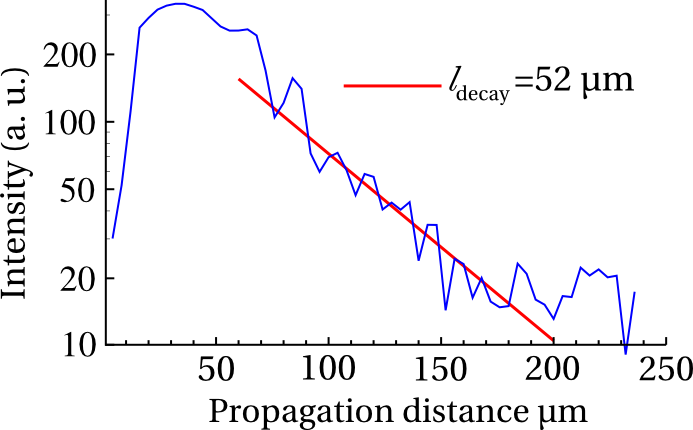}
    \caption{Integrated intensity as a function of the propagation distance for figure 2c, plotted in a logarithmic scale. From an exponential fitting a decay distance of $52$ $\mu$m, corresponding to a lifetime of $25$ ps, is deduced.}
\end{figure}

\section{Pulse frequency multiplication}
The laser pulse frequency was quadruplicated in order to have enough signal to make the image of the propagation. This was done by using two Michelson and Morley interferometers with a delay of 6 ns (180 cm) in the first one and 3 ns (90 cm) in the second one. The main limitation in the frequency multiplication is given by the time resolution of the detection: the antibunching should always be measurable in order to guarantee that single photon emission can still be identified after quadruplicate the pulse frequency, as shown in Fig.~1b of the main text. This optimization is necessary, due to the drastic attenuation of the signal after passing through the high Q factor cavity, where the non radiative losses are dominant. The photon emission increment, together with the change to transmission configuration by etching the sample surface, enhances the signal in the EMCCD camera by a factor of approximately 16. A full experimental scheme, including the laser multiplexing is shown in Fig.~\ref{setup}. The three parts of the experiment mentioned in the main text (i) the generation of single photons, ii) the injection and propagation of single polaritons and iii) detection.) are denoted by dashed black lines in the figure.

\begin{figure}[!htb]
    \centering
    \includegraphics[width=0.7\columnwidth]{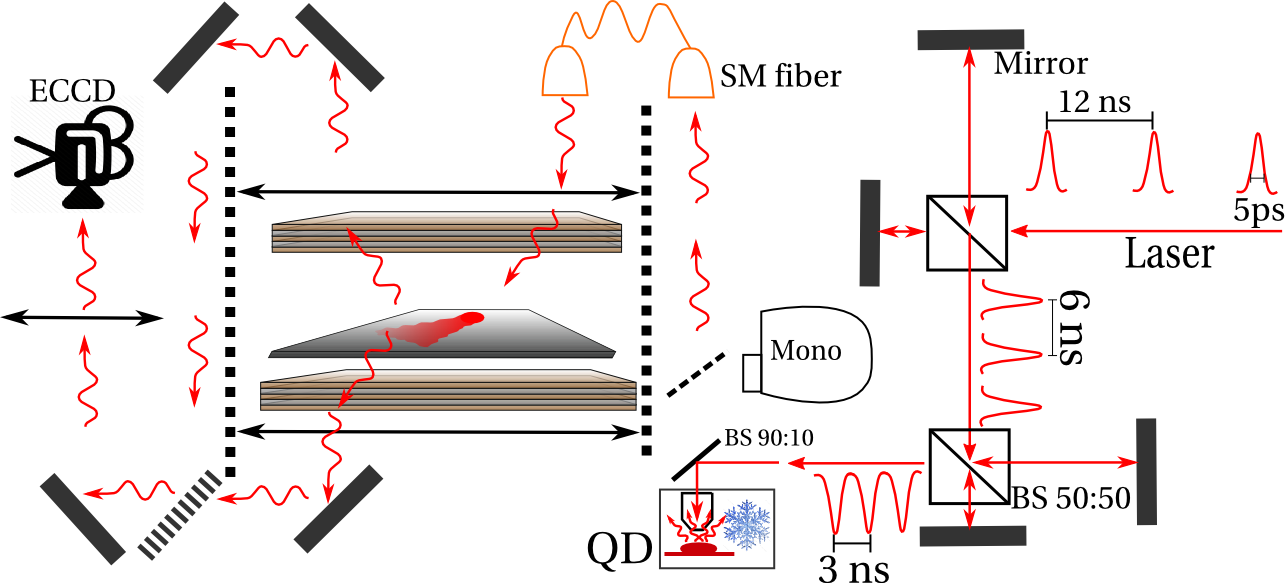}
    \caption{Schematic experimental setup including laser multiplexing, reflection and transmission configurations.}\label{setup}
\end{figure}

\section{Propagation across a finite size defect}

\begin{figure}
    \centering
    \includegraphics[width=0.5\columnwidth]{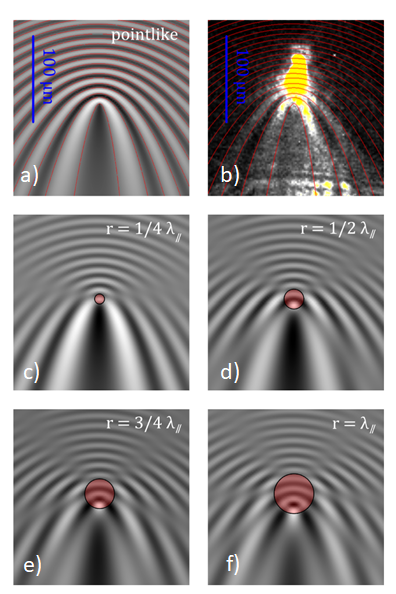}
    \caption{Interference pattern of a plane wave scattered from a defect of different radius r (highlighted by the red circle). First row: simulation (left) and experimental data (right) for the case studied in our work, i.e., a point-like defect. Second and third rows: interference pattern for a defect of increasing size going from  $r=1/4 \lambda_{//}$ to $r=\lambda_{//}$. In the case of an extended defect, higher order interferences appear. These effects are not present on our experimental data, strongly supporting our assumption of a point defect.}
    \label{size}
\end{figure}

In figure \ref{size} we show how the interference pattern is modified by the size of the defect. We compare the polariton in-plane wavelength $\lambda_{//}$ and the defect size $r$. The polariton in-plane wavelength is given by the polariton in-plane wavevector: $\lambda_{//} \approx 2\pi/k \approx 20 \mu m $, given that $k \approx 0.28 \mu m^{-1}$. In the first row, panel a) and b) show respectively the simulation and the experimental data for the case studied in this work, i.e. a point-like defect whose cross-section is much smaller than $\lambda_{//}$. In the second row, panel b) and c) show the simulation of the interference pattern for the cases $r = 1/4 \lambda$ and $r = 1/2 \lambda$. In the third row, panel e) and f) show the interference pattern for the cases $r = 3/4 \lambda$ and $r = \lambda$. These interference pattern show that when the defect radius $r$ is of the same order of magnitude as the in-plane polariton wavelength $\lambda$, a complex interference pattern arises, showing higher order interference effects, with several phase jumps across the interference fringes. These simulations clarify that the experimental fringes ahead of the defect cannot come from an overlap with the finite cross-section of the defect. In fact, the signature of a finite cross-section of the defect are not present in our experimental data. Similar results have also been shown in \cite{PhysRevLett.113.103901}.

\section{Single polariton propagation in the case of a partial scattering}
\begin{figure}
    \centering
    \includegraphics[width=0.5\columnwidth]{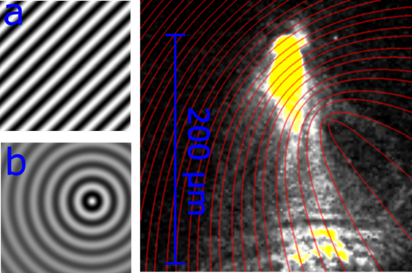}
    \caption{a) Space distribution of the electric field of an incoming plane wave with a tilted in-plane momentum but same module as the used in Fig.~4 of the main text. b) Electric field distribution for circular wave, as it could be modeled for the light scattered from a punctual structural defect in the microcavity. c) Experimental density map from the Fig. 3c. A low fraction of the incoming light reaches and scatters against the defect, and an interference pattern can be noticed which is very weak and localized close around the defect. The parabolic fronts from the model of scattering from a punctual defect fits the fringes data well, confirming the suitability of the assumption. The numerical simulations are obtained by assuming an incoming polariton wave with an in-plane momentum $k\approx 0.28μm^{-1}$, as in the experiment, but approximately tilted to $45^{\circ}$ direction.}
    \label{lateral}
\end{figure}

In this section we reproduce the pattern of the “free propagating” polariton (see figure 3c of the main text), where only a small fraction of the wave-packet is scattered by the defect, which is this time displaced around 60 μm on the right side to the central propagation axis. The wavevector direction of the incoming plane wave has been modified to account for the spreading in the polariton packet propagation, whose momentum is locally tilted under the defect, see figure \ref{lateral} a). The theoretical fitting to the barely visible interference pattern is displayed on the right side of Fig.\ref{lateral}.
The good agreement of the model with the experimental data indicates once again the suitability of the assumption of punctual structureless defect. The specific tilted direction of the momentum is an approximation which is strictly valid only locally, close around and after the defect along its tilted line of sight, which is however also the area where the intensity of the circular wave is not negligible.  

\section{Second order correlation function at different repetition rates}

\begin{figure}
    \centering
    \includegraphics[width=0.99\columnwidth]{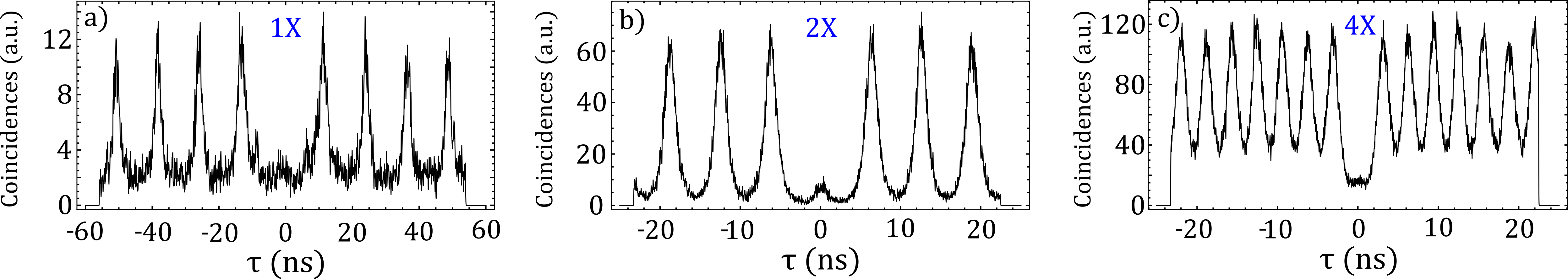}
    \caption{Second order correlation function measured before any modification of the laser repetition rate (a), after its duplication with a Michelson and Morley interferometer (b) and after its quadruplication by mean of two cascaded interferometers (c). In every case, the data confirms the regime of single photons.}
    \label{several_g}
\end{figure}

In order to increase the single photons flux arriving on the microcavity, we used a couple of Michelson and Morley interferometers. We obtained a 320 MHz pumping rate, resulting in 140 thousands single photons per second impinging on the microcavity. The low values of the second order correlation function in the three cases shown in figure \ref{several_g} confirms that the system is always in the single photon regime. 

\end{document}